\documentclass[twocolumn,prb,superscriptaddress,showpacs,amsmath,amssymb]{revtex4}
\usepackage{graphicx,dcolumn,bm}

\begin{document}

\title{Superconducting electronic state in optimally doped YBa$_2$Cu$_3$O$_{7-\delta}$ observed with
laser-excited angle-resolved photoemission spectroscopy}

\author{M.~Okawa}
\affiliation{Institute for Solid State Physics, University of Tokyo,
Kashiwa, Chiba 277-8581, Japan}

\author{K.~Ishizaka}
\affiliation{Institute for Solid State Physics, University of Tokyo,
Kashiwa, Chiba 277-8581, Japan}
\affiliation{CREST, Japan Science and Technology Agency,
Tokyo 102-0075, Japan}

\author{H.~Uchiyama}
\affiliation{Japan Synchrotron Radiation Research Institute,
Sayo, Hyogo 679-5198, Japan}

\author{H.~Tadatomo}
\affiliation{Department of Physics, Osaka University,
Toyonaka, Osaka 560-0043, Japan}

\author{T.~Masui}
\affiliation{Department of Physics, Osaka University,
Toyonaka, Osaka 560-0043, Japan}

\author{S.~Tajima}
\affiliation{Department of Physics, Osaka University,
Toyonaka, Osaka 560-0043, Japan}

\author{X.-Y.~Wang}
\affiliation{Technical Institute of Physics and Chemistry, Chinese Academy of Sciences,
Beijing 100190, China}

\author{C.-T.~Chen}
\affiliation{Technical Institute of Physics and Chemistry, Chinese Academy of Sciences,
Beijing 100190, China}

\author{S.~Watanabe}
\affiliation{Institute for Solid State Physics, University of Tokyo,
Kashiwa, Chiba 277-8581, Japan}

\author{A.~Chainani}
\affiliation{RIKEN SPring-8 Center, Sayo, Hyogo 679-5148, Japan}

\author{T.~Saitoh}
\affiliation{Department of Applied Physics, Tokyo University of Science,
Tokyo 162-8601, Japan}

\author{S.~Shin}
\affiliation{Institute for Solid State Physics, University of Tokyo,
Kashiwa, Chiba 277-8581, Japan}
\affiliation{CREST, Japan Science and Technology Agency,
Tokyo 102-0075, Japan}
\affiliation{RIKEN SPring-8 Center, Sayo, Hyogo 679-5148, Japan}

\date{\today}

\begin{abstract}
	Low energy electronic structure of optimally doped YBa$_2$Cu$_3$O$_{7-\delta}$
	is investigated using laser-excited angle-resolved photoemission spectroscopy.
	The surface state and the CuO chain band that usually overlap the CuO$_2$ plane
	derived bands are not detected, thus enabling a clear observation of the
	bulk superconducting state.
	The observed bilayer splitting of the Fermi surface is $\sim$0.08 {\AA}$^{-1}$
	along the $(0,0)$--$(\pi,\pi)$ direction, significantly larger than
	Bi$_2$Sr$_2$CaCu$_2$O$_{8+\delta}$.
	The kink structure of the band dispersion reflecting the renormalization effect
	at $\sim$60 meV shows up similarly as in other hole-doped cuprates.
	The momentum-dependence of the superconducting gap shows $d_{x^2-y^2}$-wave
	like amplitude, but exhibits a nonzero minimum of $\sim$12 meV along the $(0,0)$--$(\pi,\pi)$
	direction.
	Possible origins of such an unexpected ``nodeless'' gap behavior
	are discussed.
\end{abstract}


\maketitle

\section{Introduction}
	The high-$T_c$ superconductivity in cuprates is still attracting strong attention
	after extensive studies over twenty years.
	In particular, the concomitant technical progress of angle-resolved photoemission
	spectroscopy (ARPES) has significantly contributed to the investigation of
	high-$T_c$ cuprates since its discovery, e.g., Fermi surface determination and confirmation
	of the $d_{x^2-y^2}$-wave superconducting gap symmetry. \cite{Damascelli03}
	More recently, the ``kink'' like structure in band dispersions was observed in a number of
	hole-doped cuprates, thus shedding light on the details of the electron-boson coupling and
	the renormalization effect.
	\cite{Damascelli03,Lanzara01,Sato03,Cuk04,Sato07,Wei08}
	Moreover, very recent arguments on the pseudogap and so-called the ``Fermi arc'' in relation
	to the mechanism of superconductivity have further enhanced the significance and
	effectiveness of ARPES studies on cuprates. \cite{Tanaka06,Kondo07b,Lee07,Kanigel06,Kanigel07}
	The ARPES results reported thus far, however, are mostly dominated by the data from
	a bismuth-based cuprate Bi$_2$Sr$_2$CaCu$_2$O$_{8+\delta}$ (BSCCO) and its related compounds,
	because of the easiness of getting a clean surface by cleaving.
	For deeper understanding of the high-$T_c$ superconductivity, systematic elucidation
	of the superconducting electronic structure across various compounds is still
	highly desired.
	
	YBa$_2$Cu$_3$O$_{7-\delta}$ (YBCO) is one such typical cuprate system with
	relatively high-$T_c$ of 93 K, whose orthorhombic crystal structure is characterized by the
	existence of one-dimensional CuO chains besides bilayered CuO$_2$ planes. \cite{Jorgensen87}
	Since samples of good quality are available in YBCO, the electric, magnetic, and
	phononic properties has been extensively studied by various kinds of probes,
	thus providing a number of important findings such as pseudo-(spin) gap,
	magnetic resonance mode, etc., at a very early stage. \cite{Timusk99}
	Recently, owing to the progress of detwinning technique aligning the CuO chains
	perfectly in a single domain, in-plane anisotropic electronic properties are being
	revealed.
	Particularly, indication of the superconducting order parameter showing two-fold
	$d_{x^2-y^2} + s$-wave symmetry has been concluded by many probes including
	optical, \cite{basov95,Limonov98,Nemetschek98} phase-sensitive, \cite{Kirtley06}
	and tunneling current \cite{Smilde05} measurements.
	A precise investigation of the superconducting state in YBCO by high-resolution ARPES will be
	very valuable in order to give a comparative study across the various probes
	as well as compounds.
	A considerable amount of ARPES results have been reported so far.
	\cite{Veal94,Schabel98a,Lu01,Borisenko06,Nakayama07,Zabolotnyy07,Hossain08}
	However, a critical difficulty called the ``surface state''
	is known to hinder the bulk electronic structure study of YBCO.
	\cite{Damascelli03,Schabel98a,Lu01}
	It had been long argued that this problem should arise from two kinds of the easy cleavage,
	CuO chain and BaO terminations, as observed by scanning tunneling microscope (STM).
	\cite{Edwards92,Maki01}
	Recent effortful ARPES studies have successfully attributed the origin of
	the surface state to the first CuO$_2$ bilayer being excessively overdoped due to
	the termination. \cite{Nakayama07,Zabolotnyy07,Hossain08}
	Nakayama \textit{et al.}\cite{Nakayama07} and Zabolotnyy \textit{et al.}\cite{Zabolotnyy07}
	also attained the partial observation of the bulk superconducting state beside
	the surface state, but unfortunately, not clearly around the momentum region
	where the superconducting gap takes the minimum (nodal region).
	To discuss the total low energy electronic structure, more precise data without
	the influence of the surface state is necessary over a wide region of the
	Brillouin zone.
	
	In this article, we report the direct observation of the superconducting electronic
	state in optimally doped YBCO using laser-ARPES. By using a vacuum ultraviolet (VUV)
	laser as the light source, the ARPES measurement with sub-meV resolution and bulk
	sensitivity is available. \cite{Kiss08}
	We found clear band dispersions from the bulk state, showing the evolution of
	$d_{x^2-y^2}$ like superconducting gap.
	The surface state as well as the CuO chain band was not observed throughout
	the momentum region of measurements, thus making the analysis of the superconducting
	CuO$_2$ band much easier.
	The observed bilayer splitting is larger than BSCCO, but as expected by band calculations,
	\cite{Andersen94,Andersen95} while the kink structure at $\sim$60 meV appears similarly
	as in other hole-doped cuprates.
	The momentum-dependence of the gap in detail takes a minimum of $\sim$12 meV along
	$(0,0)$--$(\pi,\pi)$ direction, thus showing ``nodeless'' behavior.
	We discuss the possible origins of such an unusual superconducting state.

\section{Experimental Details}
	High quality single crystals of optimally doped YBa$_2$Cu$_3$O$_{7-\delta}$ ($T_c = 93$ K) 
	were grown by the crystal pulling method \cite{Yamada93} and detwinned by annealing under
	uniaxial pressure. The sample orientation was determined by x-ray Laue diffraction and Raman
	scattering measurements.
	
	Photoemission experiments were performed by using a linearly polarized laser-ARPES
	system \cite{Kiss08} equipped with a Scienta R4000 hemispherical electron analyzer and
	a quasi-continuous-wave VUV laser ($h\nu = 6.994$ eV) as a light source.
	Owing to its low kinetic energy, the escape depth of a photoelectron
	attains a value of $\sim$100 {\AA}, \cite{Seah79} thus enabling us a
	bulk sensitive measurement.
	The pressure was better than $5 \times 10^{-11}$ Torr throughout the measurements.
	The spot size of the incident light was around $\sim$200 $\mu$m.
	We obtained clean (001) surfaces of YBCO by cleaving \textit{in situ} at 5--10 K.
	Sample degradation due to aging was not observed during the continuous measurement
	over 24 h.
	We set the total energy resolution of the measurement system to 2--3 meV in order to gain
	in photoemission intensity.
	The Fermi level ($E_F$) was determined in accuracy of better than $\pm$0.5 meV
	by the Fermi edge of an evaporated Au film connected electrically to the sample.
	Unless specifically noted, the temperature of data acquisition is 5 K.
	For the transformation of angle to momentum,
	we used the value of 4.9 eV as the work function of YBCO. \cite{Rietveld92}
	For simplicity, we use ``nodal direction'' to represent the $\Gamma(0,0)$--S$(\pi , \pi)$
	direction, following the custom of a normal $d_{x^2-y^2}$
	superconductivity.

\section{Results and Discussions}	
	\subsection{Band dispersions and Fermi surface}
\begin{figure}
			\includegraphics[width=80mm]{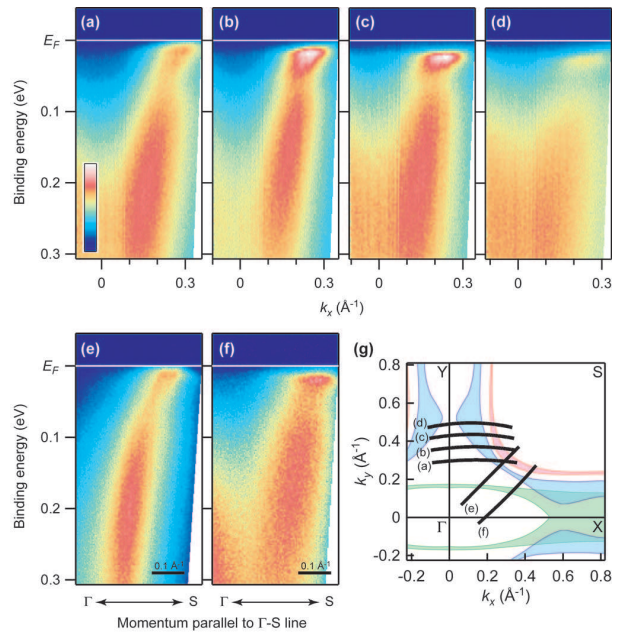}
			\caption{(Color) (a)--(f) ARPES intensity images of
				optimally doped YBCO obtained at 5 K.
				(g) Schematic Fermi surfaces of YBCO in the first Brillouin zone.
				Red, blue, and green shaded areas indicate the $k_z$-projected Fermi surfaces
				of the bonding and antibonding CuO$_2$ bands and the CuO chain band, respectively,
				obtained from a band calculation. \cite{Andersen95}
				The momentum cuts of the measurements for (a)--(f) are shown by black curves.}
			\label{fig1}
		\end{figure}

		Figures \ref{fig1}(a)--\ref{fig1}(f) show the obtained ARPES intensity images
		as a function of binding energy ($E_B$) and momentum ($k$) along momentum
		cuts as indicated in Fig.\ \ref{fig1}(g).
		Here, we can observe clear band dispersions accompanying peaks near $E_F$.
		According to the band calculations for slightly overdoped YBCO ($\delta = 0$),
		\cite{Andersen94,Andersen95}
		there are three Fermi surfaces expected; the bonding and antibonding two-dimensional
		Fermi sheets arising from the bilayer coupling of CuO$_2$ planes and the one-dimensional
		surface from a CuO chain.
		These Fermi surfaces, whose $k_z$-projections are shown by the shaded areas in
		Fig.\ \ref{fig1}(g), are expected to show large three-dimensionality ($k_z$-dependence)
		compared to BSCCO.
		The observed band dispersions in Fig.\ \ref{fig1}(a)--\ref{fig1}(f) thus correspond
		to two-dimensional CuO$_2$ plane bands.
		With approaching Y-point, the ARPES intensity becomes weak and the band dispersion
		becomes difficult to be distinguished from the background, as shown in Fig.\ \ref{fig1}(d).
		This is possibly because of the transition matrix element effect in the photoemission
		process, concerning a significant $h\nu$- and $k$-dependence in spectral weight.
		\cite{Nakayama07}
		On moving away from the nodal direction [(a, e) $\rightarrow$ (d, f) in Fig.\ \ref{fig1}],
		we can easily see that the top of the band dispersion tends to sink beneath
		$E_F$, indicative of a gap opening.
		This gapped feature clearly reveals the $d_{x^2-y^2}$-wave like magnitude, i.e.,
		the gap size $|\Delta|$ increases with approaching the antinodal (X- or Y-point) side. 
		This ensures that the present observation of band dispersions indeed reflects
		the superconducting bulk CuO$_2$ electronic structure, possibly owing to
		the bulk sensitivity and/or the matrix element effect at $h\nu = 6.994$ eV.
		The details of the superconducting gap and the coherence peak shape
		will be discussed later (section III-B).
		
\begin{figure}
			\includegraphics[width=80mm]{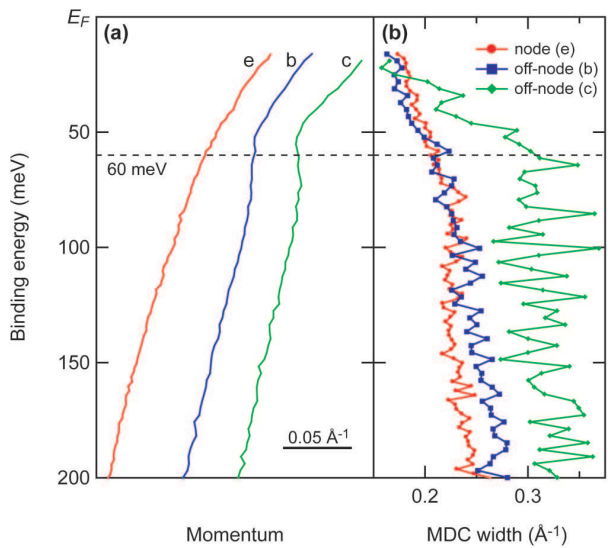}
			\caption{(Color online) (a) Band dispersions obtained from MDC fittings at nodal and off-nodal
				regions.
				Labels (e, b, and c) correspond to the momentum cuts shown in Fig.\ \ref{fig1}(g).
				(b) Peak width of MDCs as a fuction of the binding energy.
				}
			\label{fig2}
		\end{figure}

		In addition to the superconducting gap, we can also recognize a clear band
		renormalization at $\sim$60 meV, similarly as in other high-$T_c$ cuprates.
		To elucidate the band renormalization character, band dispersions along several
		momentum cuts including the nodal direction are represented in Fig.\ \ref{fig2}(a).
		These dispersions are obtained by plotting the peak positions of momentum distribution
		curves (MDCs), the momentum profiles of the ARPES intensity image at each energy.
		We estimated the peak position by fitting a Lorentzian curve to each MDC.
		We can easily find the band renormalization behavior in this dispersion at around 60 meV,
		both at nodal and off-nodal regions.
		The energy of this kink feature is quite similar to those reported previously in YBCO.
		\cite{Borisenko06,Nakayama07}
		The kink also appears as the decrease of the peak width of MDCs below $\sim$60 meV,
		as shown in Fig.\ \ref{fig2}.
		The width of MDC peak is considered to be proportional to the imaginary part of the self-energy
		$\text{Im} \, \Sigma (\omega)$. \cite{Damascelli03}
		Thus, the above result indicates the suppression of the scattering rate below the kink energy.
		These data show stronger renormalization at the antinodal side than near the nodal direction,
		which is similar to the momentum-dependence as observed in bismuth-based cuprates.
		\cite{Sato03,Cuk04}
		Thus we confirmed that the kink structure in hole-doped high-$T_c$ cuprates
		shows a universal character, also in the superconducting electronic state in YBCO.
		
\begin{figure}
			\includegraphics[width=80mm]{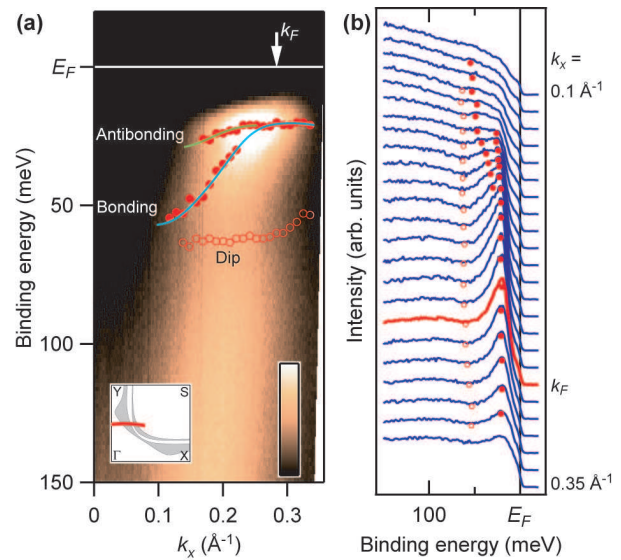}
			\caption{(Color online) ARPES intensity image (a) and EDCs (b)
				at the off-nodal location as shown in the inset of (a).
				Closed circles denote the peak positions of EDCs,
				while the dip positions are represented by open circles.
				Curves are guides for eyes to show the bonding and antibonding band
				components.
				The $k_F$ position of the outer bonding band is shown by an arrow.}
			\label{fig3}
		\end{figure}

		Since the band calculations \cite{Andersen94,Andersen95} predicted a large
		bilayer splitting of the CuO$_2$ bands, we look into the details of the
		band dispersions to see whether they are composed of two components, bonding and
		antibonding bands.
		Figures\ \ref{fig3}(a) and \ref{fig3}(b) show the band dispersion near $E_F$ at the off-nodal
		region and its energy distribution curves (EDCs) which are the energy profiles of an ARPES
		intensity image.
		At $E_B < 40$ meV, which is sufficiently below the kink energy, the band dispersion
		shows fairly sharp peaks ($\sim$20 meV) with an indication of the superconducting gap
		opening as seen in the EDCs.
		To elucidate the fine details of the band structure, we plotted the peak positions
		of EDCs shown as closed circles in Fig.\ \ref{fig3}.
		Here, we found that there are two EDC peaks at around $k_x \sim 0.2$ {\AA}$^{-1}$,
		indicating the existence of antibonding and bonding bands, as shown in Fig.\ \ref{fig3}(a).
		The Fermi wavevector ($k_F$) for the outer Fermi surface composed of bonding band
		can be determined as the $k$-point where the peak position of
		the EDC most nearly approaches $E_F$, considering the manifestation of the
		superconducting Bogoliubov quasiparticle dispersion. \cite{Norman98}
		The $k_F$ of the inner antibonding band, on the other hand, is very difficult to be
		precisely assigned, since it merges with bonding band on approaching $E_F$.
		Note that the dip structure of EDCs indicated by open circles in Fig.\ \ref{fig3}
		shows up at around 60 meV beside the quasiparticle peak, reflecting the band renormalization
		effect in the superconducting state.
		Regarding the renormalization energy at this off-nodal region, the above value
		is slightly lower compared to the 70-meV kink observed in BSCCO, \cite{Cuk04}
		and similar that in YBCO reported by Nakayama \textit{et al}. \cite{Nakayama07}
		
\begin{figure}
			\includegraphics[width=80mm]{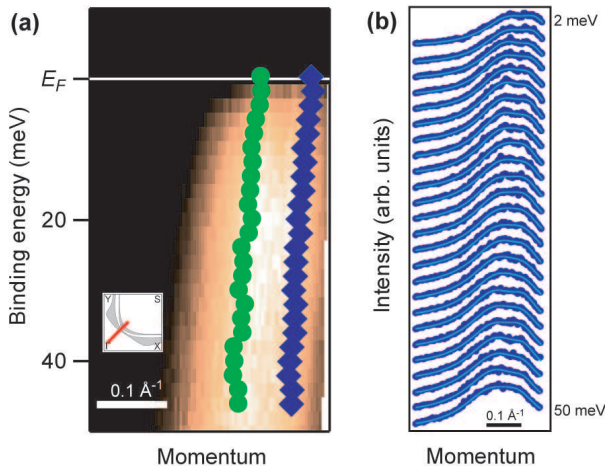}
			\caption{(Color online) ARPES intensity image (a) and MDCs (b) along the cut
				near the $\Gamma$--S direction as shown in the inset.
				Circles and squares denote peak positions of the two Lorentzians obtained by
				fitting MDCs as shown in (b), respectively indicating antibonding
				and bonding band.}
			\label{fig4}
		\end{figure}

		To separately estimate $k_F$ for respective bands, we show the ARPES spectrum
		obtained along the $\Gamma$--S direction in Fig.\ \ref{fig4}(a).
		Its MDCs are also shown in Fig.\ \ref{fig4}(b).
		We can see that each MDC contains two broad peaks.
		Each MDC was well fitted by a sum of two Lorentzians with a constant background
		$I(k) = \sum_{i=1}^2(A_i/\pi)(\Gamma_i/[(k-k_i)^2+\Gamma_i^2]) + C$
		at the energy region of $E_B < 50$ meV, as shown in Fig.\ \ref{fig4}(b).
		The peak positions obtained by the MDC fitting ($k_1$ and $k_2$) as observed in
		Fig.\ \ref{fig4}(a) suggest the bilayer splitting of about $\Delta k = 0.08$ {\AA}$^{-1}$,
		which agrees well with band calculations. \cite{Andersen94,Andersen95}
		The values of the half width at half maximum (HWHM) for respective peaks 
		($\Gamma_1$ and $\Gamma_2$), on the other hand, are about 0.1 and 0.2 {\AA}$^{-1}$ for
		bonding and antibonding bands at near $E_F$,
		which are considerably broader than those in BSCCO (e.g., 0.0031 {\AA}$^{-1}$ for
		an optimally doped sample in Ref.\ \onlinecite{Ishizaka08}).
		This broadening makes it difficult to clearly distinguish the bilayer splitting in YBCO
		despite its large value compared to BSCCO (shown later).
		It may be partly due to the greater $k_z$ dispersion expected in YBCO, giving rise to 
		broadening effect. \cite{Bansil05,Markiewicz05,Sahrakorpi05}
		To more clearly separate the bilayer split dispersion, investigations of more heavily-doped
		samples may be useful as reported in BSCCO system, which remains to be elucidated in future.
		
\begin{figure}
			\includegraphics[width=40mm]{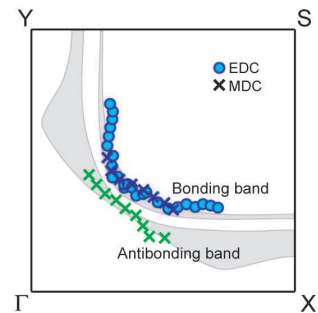}
			\caption{(Color online) $k_F$ positions plotted on the Brillouin zone.
				Circles shows $k_F$ determined as the position where EDC peak most nearly
				approaches $E_F$ (see Fig.\ \ref{fig3}).
				Crosses represent the MDC peak positions near $E_F$ (see Fig.\ \ref{fig4}).}
			\label{fig5}
		\end{figure}

		By plotting $k_F$ positions on the Brillouin zone, we can obtain the Fermi
		surface as shown in Fig.\ \ref{fig5}.
		Circles indicate the $k_F$ which were obtained by tracing the EDC peak positions
		as previously described, while the cross markers represent the MDC peak positions at $E_F$.
		We could not observe any spectra corresponding to the CuO chain band as shown
		in Figs.\ \ref{fig1}(e) and \ref{fig1}(f), which may be due to the matrix element effect.
		\cite{footnote1}
		Apparently, $k_F$ obtained from EDC is in a good correspondence with the outer
		bonding band Fermi surface we derived from MDC peaks.
		Here, we can confirm that the bilayer splitting occurs in a wide momentum region
		in YBCO, including the $(0,0)$--$(\pi , \pi)$ direction, where in BSCCO a very small splitting
		($\Delta k = 0.014$--0.015 {\AA}$^{-1}$ in Ref.\ \onlinecite{Kordyuk04} and
		$\Delta k = 0.0075$ {\AA}$^{-1}$ in Ref.\ \onlinecite{Yamasaki07}) is observed.
		Both of the bonding and antibonding band Fermi surfaces agree quite well with the
		LDA band calculation \cite{Andersen94,Andersen95} and a previous report of the nodal bilayer
		splitting in YBCO. \cite{Borisenko06}
		A past ARPES study reporting the bulk electronic state of YBCO using 46-eV photon,
		\cite{Nakayama07} nevertheless, exhibited very small bilayer splitting of
		$\Delta k \sim 0.02$--0.03 {\AA}$^{-1}$ at the off-nodal region, while our data show
		$\Delta k \sim 0.09$ {\AA}$^{-1}$
		at a similar momentum region.
		This difference indicates that the bulk electronic state of YBCO may have
		significant three-dimensionality compared to BSCCO, thus causing a strong
		$h\nu$-dependence in ARPES result. \cite{footnote2}
		
	\subsection{Superconducting gap anisotropy}
\begin{figure}
			\includegraphics[width=80mm]{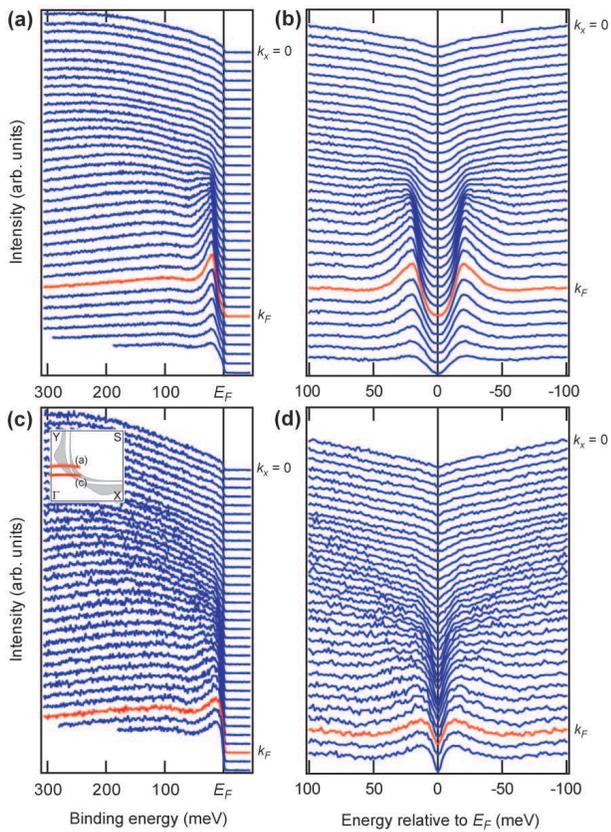}
			\caption{(Color online) (a),(c) EDCs along the momentum cut off and on
				the nodal point.
				The inset in (c) shows each momentum cut of the measurements.
				(b),(d) EDCs from (a) and (c) symmetrized at $E_F$.}
			\label{fig6}
		\end{figure}

\begin{figure}[b]
			\includegraphics[width=35mm]{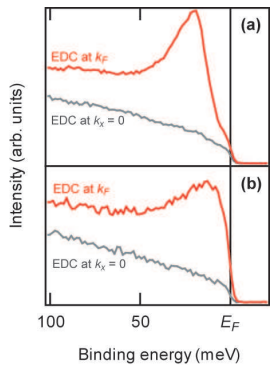}
			\caption{(Color online) (a),(b) Comparison of EDC at $k_F$ to
				that at $k_x = 0$ along off- and on-nodal momentum cuts, extracted from
				the data in Figs.\ \ref{fig6}(a) and \ref{fig6}(c), respectively.}
			\label{fig7}
		\end{figure}

		Here we focus on the evaluation of the superconducting gap and its anisotropy.
		In order to estimate the gap magnitude carefully, we show EDCs along the off-nodal
		direction and those symmetrized at $E_F$ in Figs.\ \ref{fig6}(a)
		and \ref{fig6}(b), respectively.
		The symmetrized EDC spectrum can be expressed by
		$I_{\text{sym}}(\omega) = I(-\omega) + I(\omega)$, where $I(\omega)$ is the intensity
		of EDC as a function of $\omega$, energy relative to $E_F$.
		As we described in section III-A, $k_F$ of the outer bonding band can be determined
		as the $k$-position where EDC peak most nearly approaches $E_F$, i.e.,
		the peaks forming a pair in $I_{\text{sym}}(\omega)$ get closest to each other.
		In Fig.\ \ref{fig7}(a), we show the close up of the EDCs at $k_F$ and $k_x = 0$
		extracted from Fig.\ \ref{fig6}(a).
		The EDC at $k_F$ clearly indicates a sharp peak at $\sim$20 meV and reduced intensity
		around $E_F$.
		Such spectral features represent the opening of the superconducting gap and the
		evolution of the Bogoliubov quasiparticle peak.
		The small edge-like component at $E_F$ partly arises from the (angle-integrated type)
		background, as also observed at $k_x = 0$ where no band dispersion exists near $E_F$.
		The remaining may be due to the tail-like states of the coherence peak
		with finite lifetime, while we cannot completely rule out the subtle inclusion of
		the surface state.
		Nevertheless, the dominant spectral shape is very close
		to what is observed in BSCCO, indicating that the superconducting gap can be
		precisely investigated similarly in YBCO.
		
\begin{figure*}
			\includegraphics[width=140mm]{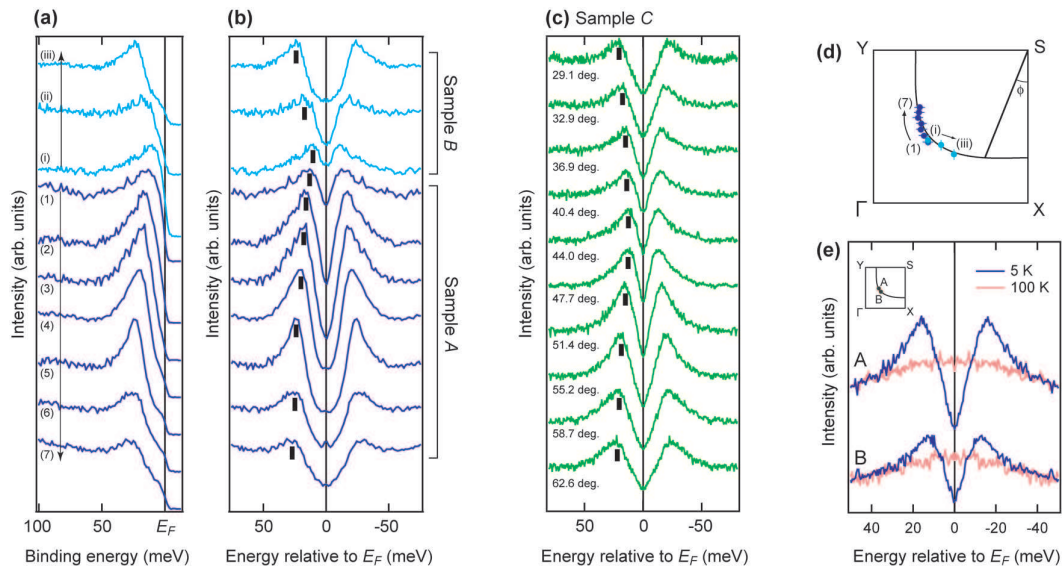}
			\caption{(Color online)
				EDCs at $k_F$ (a) and those symmetrized
				at $E_F$ (b) obtained at momentum positions
				(1)--(7) and(i)--(iii) as indicated in (d).
				(1)--(7) and (i)--(iii) are the results from two different samples (samples \textit{A}
				and \textit{B}) along momentum cuts parallel to the $\Gamma$--X and
				$\Gamma$--Y directions, respectively.
				(c) Symmetrized EDCs at $k_F$ obtained from one sample (sample \textit{C})
				along momentum cuts parallel to the $\Gamma$--S direction.
				Each $k_F$ position is represented by the Fermi surface angle $\phi$ as
				indicated in (d).
				(e) Symmetrized EDCs at the temperature of
				5 K and 100 K.
				A and B correspond to spectra at off-nodal and nodal points as shown
				in the inset.}
			\label{fig8}
		\end{figure*}

		To discuss the momentum-dependence of the superconducting gap, we show
		in Figs.\ \ref{fig8}(a)--\ref{fig8}(c) EDCs at $k_F$ and those symmetrized at $E_F$,
		for respective positions along the Fermi surface contour as shown in Fig.\ \ref{fig8}(d).
		We can see a systematic behavior that the gap size increases on approaching
		the antinodal X- and Y-points, in accord with a $d_{x^2-y^2}$-wave like
		gap symmetry.
		A surprising result appears near the nodal point, nevertheless,
		as the persistent existence of a small gap on crossing the $\Gamma$--S line.
		Even at the momentum point along $\Gamma$--S direction where the gap size becomes
		minimum [corresponds to (1) and (i) in Fig.\ \ref{fig8}(b) and $\phi \approx 45^{\circ}$
		in Fig.\ \ref{fig8}(c)], the spectrum clearly shows a gapped feature thus indicating the
		``nodeless'' character of the superconducting gap.
		Near the antinodal X and Y regions, the EDC shows a fairly weak coherence peak
		with high background level, reflecting the loss of ARPES intensity due to the matrix
		element effect.
		On approaching the $\Gamma$--S direction, on the other hand, the coherence
		peak intensity evolves and gets sharp in width, as is also observed in BSCCO.
		In the very vicinity of the $\Gamma$--S line, however,
		the coherence peak seems to become suppressed and broadened, even though the higher
		energy part ($\gtrsim$0.1 eV) of the dispersion remains clearly [Figs.\ \ref{fig1}(a) and
		\ref{fig1}(e)].
		Such behavior is fairly unusual among hole-doped high-$T_c$ cuprates, where
		the nodal quasiparticle tends to show a robust coherence.
		
		To show the validity of our data analysis, we plot the raw EDCs along the cut
		where the gap minimum appears in Fig.\ \ref{fig6}(c).
		Comparing to the off-nodal point as shown in Fig.\ \ref{fig6}(a),
		the band dispersion approaches somewhat closer to $E_F$.
		The EDCs and those symmetrized [Fig.\ \ref{fig6}(d)], however, persistently
		show a clear gap across $k_F$.
		In Fig.\ \ref{fig7}(b), we show the EDC spectrum at $k_F$ together with
		that at $k_x = 0$.
		We can clearly see a peak structure at $\sim$12 meV for EDC at $k_F$,
		which is responsible for the gap feature.
		By comparing these two spectra at $k_F$ and $k_x = 0$, we can rule out
		the possibility that this gap structure is due to the background
		(angle-integrated) component.
		In addition, we show the temperature-dependence of the symmetrized EDC
		at $k_F$ for off-nodal (A) and nodal (B) points in Fig.\ \ref{fig8}(e).
		Both of them clearly show the gap structure disappearing between 5 K and 100 K
		($> T_c = 93$ K).
		This indicates that this ``nodal gap'' evolution also reflects some
		modification of the electronic state, like superconducting transition.
		
\begin{figure}[b]
			\includegraphics[width=60mm]{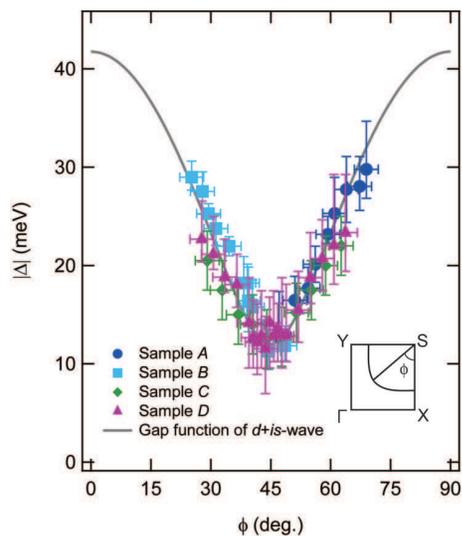}
			\caption{(Color online) The gap size $|\Delta|$ plotted as a
				function of the Fermi surface angle $\phi$ obtained from four
				different samples \textit{A}--\textit{D}.
				Here, the samples \textit{A}--\textit{C} are indical to those
				shown in Figs.\ \ref{fig8}(a)--\ref{fig8}(c).
				The curve is the gap function of
				$d_{x^2-y^2}+is$-wave, where $\Delta_d = 40$ meV and
				$\Delta_s = 10$ meV.
				The error bars arise from the experimental uncertainty of
				the EDC peak positions and the estimation of the momentum
				value.}
			\label{fig9}
		\end{figure}

		Now we discuss the gap symmetry by plotting $|\Delta|$ as a function of
		the Fermi surface angle, $\phi$ (Fig.\ \ref{fig9}).
		We estimated the gap magnitudes by taking the energy position of the
		coherence peaks in the symmetrized EDCs at $k_F$.
		In order to confirm the reproducibility, we obtained ARPES data from seven different
		samples and found that all of their results coincide within errors.
		Four of them are shown here (samples \textit{A}--\textit{D}).
		Our result indicates that the momentum-dependence of the gap in the observed region
		is nearly symmetric with respect to $\phi = 45^{\circ}$, where the gap shows
		the minimum of $\sim$12 meV.
		Note that the nodes should shift to $\phi = 45^{\circ} + \delta\phi$
		(e.g., $\delta\phi \sim -3^{\circ}$ from Ref.\ \onlinecite{Kirtley06})
		in case of $d_{x^2-y^2}+s$-wave, but do not disappear unless
		the $s$-wave component dominates that of $d_{x^2-y^2}$-wave.
		A clear difference of the gap size between X and Y regions
		(e.g., the X-Y anisotropy ratio of the gap magnitude $\Delta_{\text{Y}}/\Delta_{\text{X}}$
		is larger than $\sim$1.22 in Ref.\ \onlinecite{Smilde05})
		indicative of the $d_{x^2-y^2}+s$-wave like anisotropy, is also not observed.
		Precise measurements covering the whole Brillouin zone (both X- and Y-points)
		are further necessary, nevertheless, to make any definite conclusion on the X-Y
		anisotropy of the gap. \cite{footnote1}
		Assuming that the observed gap character arises from a single superconducting state,
		a time reversal symmetry broken state such as $d_{x^2-y^2}+is$- or
		$d_{x^2-y^2}+id_{xy}$-wave must be taken into account in order to reproduce such
		a nodeless $d_{x^2-y^2}$-wave like symmetry. \cite{VanHarlingen95,Tsuei00}
		Here we tentatively tried fitting the $\phi$-dependence with the $d_{x^2-y^2}+is$
		symmetry gap function $\Delta_{d+is} = \Delta_d \cos (2\phi) + i\Delta_s$
		as shown in Fig.\ \ref{fig9},
		and obtained $\Delta_d = 40$ meV and $\Delta_s = 10$ meV.
		
\begin{figure}
			\includegraphics[width=80mm]{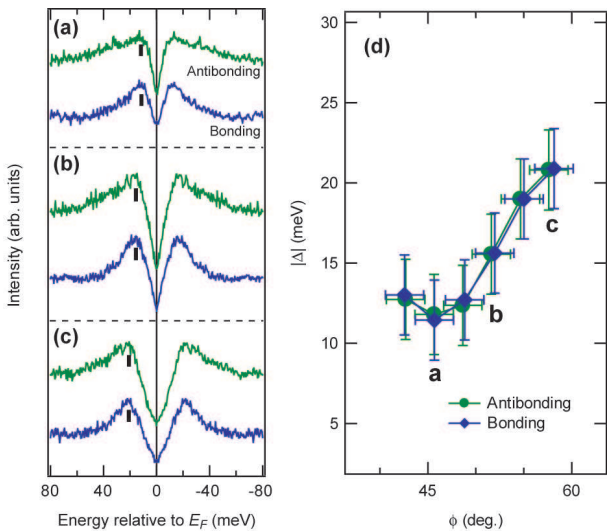}
			\caption{(Color online) (a)--(c) Symmetrized EDCs at $k_F$ along antibonding
				and the bonding band Fermi sheets.
				Estimated gaps for respective bands are shown in (d)
				as a function of $\phi$.}
			\label{fig10}
		\end{figure}

		The Fermi surface-dependences of the superconducting gap magnitude and its
		anisotropy should be also considered, which are actually observed in
		several multiband superconductors such as
		$2H$-NbSe$_2$, \cite{Yokoya01,Kiss07}
		MgB$_2$, \cite{Souma03,Tsuda03}
		and a novel FeAs-based high-$T_c$ superconductor. \cite{Ding08}
		In Fig.\ \ref{fig10}, we show the $\phi$-dependence of $|\Delta|$ for the
		two bilayer-split bonding and antibonding bands.
		The symmetrized EDCs at $k_F$ are shown along respective Fermi sheets, where
		each $k_F$ was estimated from the MDC fitting as shown in Fig.\ \ref{fig4}.
		$\phi$-dependence of $|\Delta|$ for bonding and antibonding bands are similar
		to each other, both showing the gap minimum of $|\Delta| \sim 12$ meV at
		around $\phi = 45^{\circ}$ as discussed above.
		From this result, we cannot find any significant Fermi sheet-dependence
		of $|\Delta|$ at least in this momentum region of measurement.
		
		Now let us discuss the possible origins of the ``nodeless'' behavior of
		the superconducting gap.
		This result is in a strong contrast to many recent macroscopic experimental reports on
		YBCO.\cite{Limonov98,Nemetschek98,Smilde05,Kirtley06}
		Such behaviour has not been either observed by ARPES in other cuprate superconductors
		until now.
		To explain this unusual data, here we discuss its possible origins,
		mainly in comparison with BSCCO.
		The first possibility we must suggest is the possible effect of the surface roughness
		in YBCO arising from two types of the surface terminations (CuO chain and BaO layers)
		obtained upon cleaving. \cite{Edwards92,Maki01}
		If there is a serious microscopic roughness at the surface,
		it may disturb the conservation law of the photoelectron momentum as the boundary
		condition.
		Thus the ``nodal gap'' might be reflecting the average gap value around
		the nodal point in the momentum space.
		If this is the case, the spectrum at off-nodal region should show a broader
		peak feature than that at node, because of the greater gap distribution
		as a function of $\phi$.
		The EDC spectrum at off-nodal point, nevertheless, actually shows a much sharper
		feature as shown in Fig.\ \ref{fig7}, which does not agree well with the above
		scenario.
		Another major difference between YBCO and BSCCO
		is the existence of the CuO chain structure along $b$-axis,
		which may cause $D_{2h}$ symmetry-like modification of the gap as represented by
		the $d_{x^2-y^2}+s$-wave.
		It is unlikely to be the origin of $d_{x^2-y^2}+is$-like superconducting gap
		as we observed, however, which holds $D_{4h}$ symmetry.
		The remaining factors we should consider are the higher out-of-plane electron hopping
		and the greater bilayer coupling compared to BSCCO and related materials,
		which generate two well separated bonding and antibonding Fermi sheets with
		relatively high three-dimensional dispersions.
		Note that the ratio of out-of-plane and in-plane resistivity is
		$\rho_c/\rho_{ab} \sim 10^2$ in YBCO, while $\rho_c/\rho_{ab} \sim 10^5$
		in BSCCO. \cite{Ott03}
		Such three-dimensional electronic structure, which is not usually taken into account,
		may be playing a relevant role.
		For example, some unexpected electronic state at the (001) surface may show up,
		as the time reversal symmetry broken superconducting state discussed
		in the (110) surface of a $d_{x^2-y^2}$-wave superconductor.
		\cite{Matsumoto95a,Covington97}
		In addition, the strong three-dimensional dispersion can induce the electron pairing
		along the $k_z$ direction.
		A muon spin rotation ($\mu$SR) study \cite{Khasanov07b} actually
		revealed the possible existence of an $s$-wave superconducting component
		along $c$-axis.
		Theoretical consideration how a $k_z$-dependent superconducting component
		can be detected by various kinds of probes, including ARPES, is needed.
		Finally, we refer to several interesting reports regarding the pseudogap
		behavior in under and optimally doped YBCO. \cite{Sonier01,Xia08}
		These studies insist that a hidden order phase, such as charge/spin
		density waves \cite{Kivelson03} or time reversal symmetry broken staggered
		current flux states, \cite{Varma97,Chakravarty01} coexists with
		the superconducting phase below the pseudogap temperature ($T^{\ast} = 40$--60 K
		in optimally doped YBCO).
		This pseudogap phase may be associated with the nodal gap evolution.
		The origins cannot yet be clarified at present, nevertheless, it shows a possibility
		of a novel unconventional superconducting state realizing in YBCO.
		Further precise experimental and theoretical studies for its microscopic
		description are highly desired.
		
\section{Conclusions}
	We have investigated the low energy electronic structure of optimally-doped YBCO
	by using VUV laser-ARPES.
	The unwanted surface state as well as CuO chain band were not detected perhaps because of
	the bulk sensitivity or the matrix element effect at $h\nu = 6.994$ eV.
	Thus we succeeded in the clear observation of the superconducting state in YBCO,
	which had been long desired.
	As a universal feature for high-$T_c$ cuprates, the band renormalization effect
	caused by the electron-boson coupling was found at $\sim$60 meV.
	The wide bilayer splitting of the Fermi surface was also confirmed
	for the superconducting electronic state, in good accordance with band
	calculations.
	Regarding the momentum-dependence of the superconducting gap, we found the
	$d_{x^2-y^2}$ like behavior but without a clear evidence of gap node, showing the
	gap minimum of $\sim$12 meV along the $\Gamma$--S direction.
	This gap feature was similarly observed for both of the bilayer-split Fermi surfaces.
	This ``nodeless'' character of the gap raises the possibility of an unusual
	superconducting state appearing in YBCO, thus motivating further
	systematic studies.
	
\begin{acknowledgments}
	The authors thank M.\ Ogata, T.\ Tohyama, and T.\ Yokoya for fruitful
	discussions, and T.\ Kiss, T.\ Shimojima, and S. Tsuda for technical help.
	This work was partially supported by Grants-in-Aid for Scientific Research from the Japan Society
	for the Promotion of Science and the Ministry of Education, Culture, Sports, Science
	and Technology (MEXT), Japan.
	M.O.\ acknowledges financial support from the Global COE Program
	``the Physical Science Frontier,'' MEXT, Japan.
\end{acknowledgments}

\bibliographystyle{apsrev}
\bibliography{references}

\begin{thebibliography}{58}
\bibitem[{foo()}]{footnote1}
\bibinfo{note}{In all the samples which we succeeded in observing the superconducting
	dispersion, CuO chain band was not detected.
	In the low-energy photoemission process using 6.994-eV light, the final state is
	considered to be a rather band like character, unlike the free electron picture assumed
	to be held in conventional high-energy photoemission spectroscopy.
	This will make the matrix element effect strongly dependent on the incident light
	energy in relation to the final state, and make it difficult to quantitative simulate its
	effect.
	The $h\nu$-dependent ARPES using a low-energy synchrotron light will be useful
	to obtain more information concerning the matrix element effect
	in YBCO.}
\end{thebibliography}

\end{document}